\pdfoutput=1
\documentclass[twocolumn]{aastex63} 

\usepackage{graphicx}
\usepackage{epstopdf}	
\usepackage{booktabs}
\usepackage{lineno}

\usepackage{soul}
\usepackage{amsmath,amssymb}
\usepackage{textcomp}
\usepackage{gensymb}
\usepackage{CJK}

\shorttitle{LSST Fringing Simulation}
\shortauthors{Guo et al.}
\newcommand{\ujyac}{$\mu\rm{Jy}\ \rm{arcsec}^{-2}$}
\newcommand*\chem[1]{\ensuremath{\mathrm{#1}}}
\graphicspath{{./}{figure/}}

\submitjournal{PASP}

\begin{document}

\begin{CJK*}{UTF8}{gbsn}
\title{Fringing Analysis and Simulation for the Vera C. Rubin Observatory's Legacy Survey of Space and Time} 

\correspondingauthor{Zhiyuan Guo}
\email{zhiyuan.guo@duke.edu}

\author[0000-0001-9557-9171]{Zhiyuan Guo(郭\ 致远)}
\affiliation{Department of Physics, Duke University, Durham NC 27708, USA}

\author[0000-0003-2035-2380]{C. W. Walter}
\affiliation{Department of Physics, Duke University, Durham NC 27708, USA}

\author[0000-0002-9601-345X]{Craig Lage}
\affiliation{Department of Physics, University of California, Davis, 1 Shields Avenue, Davis CA, 95616, USA}

\author[0000-0003-1666-0962]{Robert H. Lupton}
\affiliation{Princeton University, Princeton, NJ, U.S.A.}

\author{The LSST Dark Energy Science Collaboration}

\begin{abstract}
The presence of fringing in astronomical CCD images will impact photometric quality and measurements. Yet its impact on the Vera C. Rubin Observatory's Legacy Survey of Space and Time (LSST) has not been fully studied. We present a detailed study on fringing for Charge-Coupled Devices (CCDs) already implemented on the Rubin Observatory LSST Camera's focal plane. After making physical measurements and knowing the compositions, we have developed a model for the e2v CCDs. We present a method to fit for the internal height variation of the epoxy layer within the sensors based on fringing measurements in a laboratory setting. This method is generic enough that it can be easily modified to work for other CCDs. Using the derived fringing model, we successfully reproduce comparable fringing amplitudes that match the observed levels in images taken by existing telescopes with different optical designs. This model is then used to forecast the expected level of fringing in a single LSST y-band sky background exposure with Rubin telescope optics in the presence of a realistic time varying sky spectrum. The predicted fringing amplitude in LSST images ranges from $0.04\%$ to $0.2\%$ depending on the location of a CCD on the focal plane. We find that the predicted variation in surface brightness caused by fringing in LSST y-band skybackground images is about $0.6$ \ujyac, which is 40 times larger than the current measurement error. We conclude that it is necessary to include fringing correction in the Rubin's LSST image processing pipeline.

\end{abstract}

\keywords{CCD, Sensor Anomaly, Fringing}

\section{Introduction}
Fringing in astronomical Charge-Coupled Device (CCD) detectors results from the interference of incident and reflected light between multiple layers within the CCD chip. The non-uniformity of the layers' thickness creates varying interference conditions that lead to the observed fringing pattern, which becomes prominent in the Near-Infrared (NIR) as the declining absorption coefficient of the photo-sensitive silicon layer makes the silicon more transmissive to photons. Night-sky emission lines produced by excited molecules and radicals in the upper atmosphere are the primary source of the light that causes fringing in direct imaging. The Vera C. Rubin Observatory's Legacy Survey of Space and Time (hereafter LSST) aims to explore in six optical bands (u, g, r, i, z, y) ranging from 320 to $1000\ nm$~\citep{LSST09}. The fact that non-trivial fringing patterns are observed in night images taken by a LSST prototype sensor with LSST y filter at a Naval Observatory~\citep{Brooks17} suggests that fringing may appear in images taken by Rubin Observatory's camera. It is important to account for the fringing effects properly to obtain accurate measurements, especially for LSST NIR images since observations in the NIR are crucial for transient studies such as supernovae. Thus, a deep understanding of LSST fringing patterns is needed and a fringing model needs to be added to the LSST image simulation tool~\citep{LSST21} in order to verify that the fringe removal algorithm in the LSST image processing pipeline works properly. 

The LSST focal plane array consists of 189 fully-depleted 4k$\times$4k  pixel CCD sensors made by two manufacturers, Arizona Image Technology Laboratory (ITL) and Teledyne-e2v (e2v). The photo-sensitive silicon regions of these sensors are made $100\ \mu m$ thick to improve the IR sensitivity~\citep{Connor19}. These back-illuminated CCDs are assembled into 21  Raft Tower Modules (RTM) in groups of 9 on the science focal plane~\citep{Connor16}. 

Knowing the composition of a CCD's structure, namely the material of each layer, the fringing amplitude can be deduced from geometrical optics calculation. Many efforts have been made to account for fringing in images taken by previous telescopes.~\citet{Malumuth03} devised a multi-layer fringing model and used the observed fringing amplitude across Hubble Space Telescope (HST)'s STIS CCD to derive the spatial variation of the photo-sensitive silicon layer.~\citet{Walsh03} used a similar method to fit the thickness of HST ACS WFC and HRC CCD layers. In this study, we develop a fringing model for LSST e2v sensors and fit the model to flat field data obtained from electro-optics (EO) test stands at the SLAC National Accelerator Laboratory for nine e2v CCDs in one LSST science raft using a pixel-by-pixel fitting method adopted from~\citet{Malumuth03}. Via fitting the fringing amplitude derived from flat fields, thickness map of the epoxy layer that causes the fringing observed in lab is obtained for each of the nine sensors. The fringing model is then used to forward model realistic fringing in LSST sensors based on telescope optics and sky emission lines. 

This paper is structured as follows. In Section~\ref{sec: LSST_fringe_general}, we present the fringing model and the e2v CCD structure used in this study.  This model is then used to verify the epoxy layer that gives rise to the fringing pattern observed in lab data. Fringing in ITL CCD is also discussed in this section. In Section~\ref{sec:Frining_fitting}, we describe the data reduction and fitting algorithm and present the sensor fitting results. In Section~\ref{sec:recipes for real image}, we discuss sky emission lines, LSST telescope optics and their relevance to  simulating realistic fringing in sky background image. In Section~\ref{sec:real_sky}, we verify the fringing model by applying it to simulate the fringing amplitudes of CCD sensors implemented on MonoCam and Hyper-Suprime Camera and compare those with observed values. Finally, the forward modelling of fringing in LSST sky background images is presented.

\section{LSST Fringing in General and simulation} \label{sec: LSST_fringe_general}

\subsection{Fringing Simulation in General} \label{sec:fringing model}
Based on previous studies~\citep{Malumuth03,Walsh03}, a multilayer optical model can be constructed for simulating fringing if the material composition for every layer of a CCD is known. The problem is equivalent to determining the electromagnetic solution for light wave travelling through stratified thin films, which can be solved by using the Transfer-Matrix Method. One key feature of this method is that the infinite series of interfering waves between layers are implicitly taken into account in the calculation. In this study, we use \verb|TMM|\footnote{\url{https://github.com/sbyrnes321/tmm}}, a python package developed by~\citet{Byrnes16}, to solve such problems. We only illustrate key equations here and refer readers to the reference above for more details.

If $r_{p,p+1}$ and $t_{p,p+1}$ denote the reflection and transmission going from $p^{th}$ to $p+1^{th}$ layer, ${d_p}$ as the thickness of layer $p$ and $k_p$ as the (complex) refractive index corresponding to the material in the $p$th layer, then the relations between $v_p, v_{p+1}$, the amplitude of forward travelling wave at $p^{th},p+1^{th}$ side and, $w_p,w_{p+1}$, the amplitude of backward travelling wave at $p^{th},p+1^{th}$ side, can be derived to be as :

\begin{equation*}
    \binom{v_p}{w_p} = M_n\binom{v_{p+1}}{w_{p+1}}
\end{equation*}
\begin{equation*}
    M_p \equiv \begin{pmatrix} e^{-i\delta_p} & 0 \\ 0 & e^{i\delta_p} \end{pmatrix}\begin{pmatrix} 1 & r_{p,p+1} \\ r_{p,p+1} & 1 \end{pmatrix}\frac{1}{t_{p,p+1}}
\end{equation*}
with $\delta_p = d_p k_p$.
The matrix relates the wave entering and exiting the stack via equation ($14$) in \citet{Byrnes16}:

\begin{equation*}
    \binom{1}{r} = \begin{pmatrix}\Tilde{M_{00}}&\Tilde{M_{01}}\\\Tilde{M_{10}}&\Tilde{M_{11}} \end{pmatrix}\binom{t}{0}
\end{equation*}
where

\begin{equation*}
    \tilde{M_{0,1}} = \frac{1}{t_{0,1}}\begin{pmatrix}1 & r_{0,1} \\ r_{0,1} & 1\end{pmatrix}M_1M_2\cdots M_{N-2}\ .
\end{equation*}
Thus, the transmitted, T, and reflected power, R, can be calculated as follows:

\begin{align*}
    R &= |r|^2   \\
    \mbox{s-pol}:\qquad T &= |t|^2\frac{\mbox{Re} [n\cos\theta]}{\mbox{Re}[n_0\cos \theta_0]}  \\
    \mbox{{p-pol}: } \qquad T &= |t|^2\frac{\mbox{Re} [n\cos\theta^*]}{\mbox{Re}[n_0\cos{\theta_0}^*]}
\end{align*}
where $\theta_0$ and $\theta$ are the light propagation angles in the previous and present layer calculated based on Snell's law, $n$ is the refractive index, s(p)-pol stands for s(p)-polarized light that has an electric field polarized perpendicular (parallel) to the plane of incidence, the total absorption power of the stack is given by:
\begin{equation*} 
    A = 1 - T - R\ .
\end{equation*}
Since electron-photon pairs are generated in the detection layer of CCD, we are interested in the absorption in the silicon detection layer of the stack model. This is achieved in \verb|TMM| by calculating the energy flow (Poynting vector) at the beginning of each layer.
\begin{align*}
    \mbox{s-pol:}\qquad S\cdot\hat{z} &= \frac{\mbox{Re}\left[(n)(\cos\theta)({E^*_f}+{E^*_b})(E_f-E_b)\right]}{\mbox{Re}\left[n_0 \cos\theta_0\right]} \\
    \mbox{p-pol:}\qquad S\cdot\hat{z} &= \frac{\mbox{Re}\left[(n)(\cos\theta^*)({E_f}+{E_b})(E^*_f-E^*_b)\right]}{\mbox{Re}\left[n_0 \cos\theta^*_0\right]} \\
\end{align*}
where $E_f$ and $E_b$ are the E-field for the forward and backward travelling wave at that point ($z = 0$) in the layer of interest.
Based on the assumption that all reflected light remains in the initial layer and all transmitted light get absorbed in the final layer, the power absorption in each layer can be obtained by taking the difference of energy flow calculated above between each consecutive pair of layers. The fringing model implemented in this study is performed on pixel-by-pixel level. All the calculations presented in this paper are polarization-averaged.

\begin{deluxetable*}{llll} \label{tbl:e2v Structure}
\tablecaption{Structure model of e2v-CCD250}
\tablecolumns{4}
\tablenum{1}
\tablewidth{0pt}
\tablehead{
\colhead{Layer} &
\colhead{Purpose} &
\colhead{Material} & 
\colhead{Thickness [$\mu m$]}
}
\startdata
0 ........& Ambient medium & \chem{Vacuum} & Inf. \\
1 ........& AR coating & \chem{MgF_2} \citep{Li80} & 0.1221  \\
2 ........& AR coating & \chem{Ta_2O_5} \citep{Marcos16}& 0.0441  \\
3 ........& Detection & \chem{Si} \citep{Green08} & 100  \\
4 ........& Gate Oxide & \chem{SiO_2} \citep{Malitson65} & 0.1  \\
5 ........& Gates & \chem{Si} &  0.3  \\
6 ........& Insulation & \chem{SiO_2} & 1.  \\
7 ........& Adhesive & \chem{Epoxy} & CCD dependent  \\
8 ........& Support & \chem{Si} & $165$  \\
9 ........& Substrate & \chem{Si_3N_4} \citep{Philipp73}& Inf.  \\
\hline
\enddata
\end{deluxetable*}

\subsection{e2v CCD 250} \label{sec:sensor structure}
The multi-layer optical model used to characterize the structure of e2v CCD used in this study is derived from the physical measurements from~\citet{Craig19b} as such information is not provided by the vendor. Detailed descriptions of the e2v CCD stack model and corresponding parameters are listed in Table~\ref{tbl:e2v Structure}. The top layer (layer 0) in the model is vacuum in which the photons travel before encountering the CCD. Layer 1 - 6 constitute the CCD chip structure. For the anti-reflection (AR) coating (layer 1 and 2), we use the AR coating material compositions and values derived through fitting the Quantum Efficiency (QE) measurement for a LSST CCD (Andy Rasmussen, private communication). Based on inspection on the cross section of the sensor, the e2v CCD is glued to a support silicon (layer 8) via a epoxy layer (layer 7) ~\citep{Craig19b}. The final layer (layer 9) is the substrate beneath the whole stack. The multi-layer stack model implemented in \verb|TMM| requires the ambient medium  and substrate to have infinite thickness. 

The calculations introduced in Section~\ref{sec:fringing model} are sensitive to the optical properties of the materials in the stack model. References to measured refractive indices, as a function of wavelength, of the materials implemented in the fringing model are listed in Table~\ref{tbl:e2v Structure}. The temperature dependence of silicon's refractive index and extinction coefficient is included in the model. The normalized temperature coefficients from~\citet{Green08} are used to calculate those two values at different temperatures. It is noteworthy that the measured doping level of the p-type doped silicon in LSST CCD is about $2\times10^{12} \text{cm}^{-3}$~\citep{Craig19b}. This doping level is too small to significantly impact optical properties of silicon~\citep{Jellison81}.

Previous studies~\citep{Malumuth03,Walsh03} attributed the fringing patterns observed in other CCD sensors to the spatial variations of the silicon detection layer of CCDs. However, this is not the case for fringing pattern observed in the e2v senors. For light with normal incidence, which is a good assumption for the lab data that will be discussed later, the minimum resolution of the monochromator required to resolve fringing related to a material with certain thickness at a particular wavelength 
is given by:

\begin{equation}  \label{eq:Fringe spacing}
    \delta\lambda = \frac{\lambda^2}{2nd + \lambda}
\end{equation}
where $n$ is the refractive index, $d$ is the thickness and $\lambda$ is the wavelength at which the fringing spacing is evaluated. For a $100\ \mu m$ thick silicon layer with $n_{\mathrm{si}} = 3.6$ at $\lambda = 960\ nm$, the spacing between two adjacent fringes is about $1.2\ nm$. This implies that an illumination bandwidth of less than $1.2\ nm$ is needed to resolve the fringes related to spatial variations of silicon layer. With any bandwidth greater than $1.2\ nm$, these fringes will be smeared out by the large bandpass of the illumination setup.
\begin{figure}[tb]
\centering
\includegraphics[scale = 0.45]{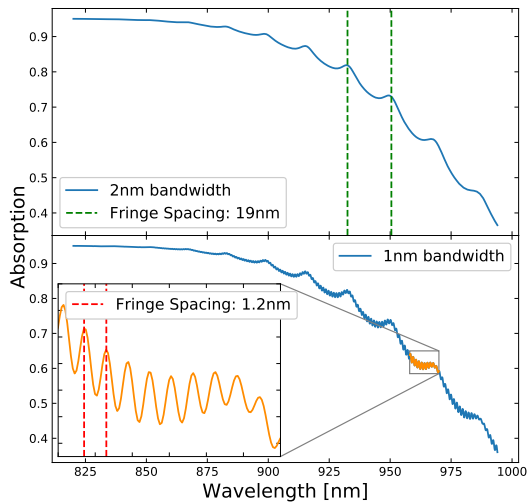}
\caption{Absorption power versus wavelength of e2v CCD based on different illumination bandwidths. {\it Upper panel:} Calculation based on $2\ nm$ bandwidth ($d_{\mathrm{epoxy}}=14\ \mu m$ ). Spacing between {\it green dashed lines:} $19\ nm$ fringe spacing for first fringing pattern related to epoxy layer. {\it Lower panel:} Calculation results based on $1\ nm$ bandwidth. Spacing between {\it red dashed lines:} $1.2\ nm$ fringe spacing for second fringing pattern related to the Si detection layer.}
\label{fig:bandwidth_sim}
\end{figure}

\begin{figure}[t]
\centering
\includegraphics[scale = 0.45]{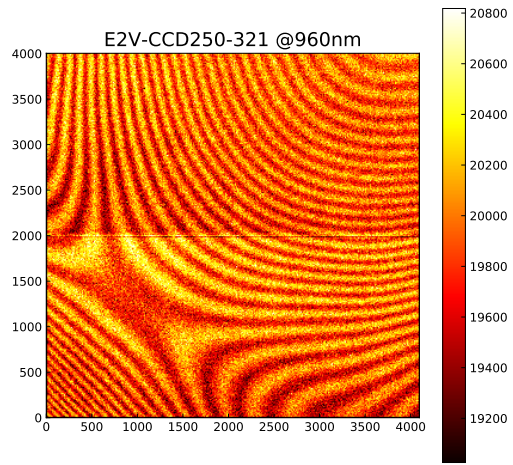}
\caption{SLAC-TS8 e2v-321 flat field taken at wavelength $\lambda = 960\ nm$. The color bar shows the number of electrons per pixel. Only fringing related to the epoxy layer is present, since the fringing caused by the non-uniformity of the silicon layer is averaged out by the large bandpass of the monochromator light.}
\label{fig:e2v_example}
\end{figure}
This argument is also supported by simulation. Based on the stack model shown in Table~\ref{tbl:e2v Structure} with a constant epoxy thickness of $14\ \mu m$, two sets of simulations with different assumptions for illumination bandwidths, $1\ nm$ and $2\ nm$, are generated. A Gaussian distribution is assumed for the monochromatic light profile in which the illumination bandwidth is the Full Width at Half Maximum (FWHM) of the Gaussian. The two panels in Figure~\ref{fig:bandwidth_sim} show absorption power for the stack as a function of wavelength for the two bandwidths respectively. The upper panel shows the calculation results for $2\ nm$ bandwidth and the calculation for $1\ nm$ bandwidth is presented in the lower panel. With a $2\ nm$ illumination bandwidth, only one fringing pattern with $19\ nm$ spacing between fringes is observed in the simulation result. Using Eq~\ref{eq:Fringe spacing}, it can be easily verified that this spacing corresponds to fringing caused by the $14\ \mu m$ epoxy layer with $n_\mathrm{epoxy} = 1.6$ at around $960\ nm$. With $1\ nm$ bandwidth, a second fringing pattern with much smaller amplitude appears, as shown in the inset figure in the lower panel. At around $960\ nm$, the second set of fringes have a spacing about $1.2\ nm$, coming from the $100\ \mu m$ silicon detection layer with monochromatic light of normal incidence.

Before being assembled into the focal plane, CCDs on each LSST science raft are sent to SLAC National Accelerator Laboratory for comprehensive tests and integration~\citep{Bond18,Ivezi19}. The data analyzed in this paper come from SLAC Test Stand 8 (henceforth SLAC-TS8). The monochromatic light used to obtain flat field data is generated from a $4"$ integrating sphere which is set $1$ meter away from the CCD. Thus, a collimated beam of light falls at normal incidence is a good approximation, which is the assumption that all the calculations in this section are based on. Figure~\ref{fig:e2v_example} shows an example of the fringing pattern observed in one particular e2v CCD taken under monochromatic light of approximately $2nm$ bandpass in SLAC-TS8. It is noticeable that only one fringing pattern is observed in this e2v CCD. This finding implies that the observed fringing should correspond to the thickness variation in the epoxy layer since any fringing related to silicon detection layer will be smeared out by the large bandpass of the illumination setup.

Another test stand setup nearly identical to SLAC-TS8 was built at Brookhaven Nation Labotory (BNL-TS8) to test LSST CCD sensors as well. BNL-TS8 has a narrower monochromator bandpass compared to SLAC-TS8. And data taken for e2v sensors in BNL-TS8 indeed shows a second set of fringing pattern, which should be related to the $100\ \mu m$ silicon based on the conclusions drawn from simulations. However, for this study, we only focus on fringing in the epoxy layer.This is because simulation results imply that the amplitude of the fringes from the epoxy layer is much greater than that from the silicon layer.  Additionally, when simulating realistic fringing in telescope image, the simple assumption of normal incident light for lab illumination setup will be replaced with telescope optics. The large aperture of telescope tends to average out the fringing patterns and further decreases the observed amplitude~\citep{Groom17}, making the impact of this second fringing pattern in silicon layer negligible compared to the one in epoxy layer. See Section~\ref{sec: ITL sensor} and Section~\ref{sec:LSST optic} for more discussion.

\subsection{ITL STA3800C CCD} \label{sec: ITL sensor}
Fringes are not observed in ITL sensors at SLAC-TS8. The reason is two-fold. First, as mentioned in Section~\ref{sec:sensor structure}, the silicon layer fringing is expected to be washed out by the large bandpass of SLAC-TS8 illumination. Second, because the ITL sensor has an additional Litho-Black coating applied to it. This highly-absorbent black coating will absorb light passing through the epoxy layer and greatly reduce the amount of reflected light from the epoxy. With illumination light coming from a monochromator with smaller bandpass compared to SLAC-TS8, fringing is observed in ITL sensors at BNL-TS8. However, study have shown that when using a $f/1.2$ beam, which closely resembles the overall focal ratio of LSST telescope ($f/1.23$)~\citep{Ivezi19}, fringing is not observed in ITL sensors even with 1nm monochromator bandpass (Lage Craig, private communication). This fact supports the argument we made in the end of Section~\ref{sec:sensor structure}. We conclude that fringing caused by the thickness variation of $100\ \mu m$ silicon layer will be trivial for LSST therefore this study focuses on the fringing pattern related to the epoxy layer in e2v sensors.

\section{Fitting thickness of Epoxy layer} \label{sec:Frining_fitting}

\subsection{SLAC Test Stand Flat Fields Data} \label{subsec:SLAC-TS8}

To simulate the observed fringing pattern from the fringing model, the thickness of the epoxy layer ($d_{\mathrm{epoxy}}$) must be derived at each pixel across the CCD. This is achieved by fitting for $d_{\mathrm{epoxy}}$ based on the observed fringing amplitude as a function of wavelength for every pixel of a CCD. Thus, the number and sampling of available data points are crucial for constraining the thickness of the epoxy layer. A series of flat fields at different wavelength were measured for e2v CCD sensors with monochromatic light illumination in SLAC-TS8 under the temperature of $-90 \degree C$. The smallest wavelength spacing between each successive flat field data available in SLAC-TS8 is $10\ nm$ for nine e2v sensors implemented on RTM-020. As it will shown in Section~\ref{sec:Fitting Results}, we are able to obtain reasonable fitting results with this sampling in wavelength.
In the following sections, we describe the method used to derive the epoxy thickness map for all of the nine CCDs.

\subsection{Data Reduction} \label{subsec: data reduction}
The flat field data are preprocessed using LSST Data Management software~\citep{Juri17,Axelrod10}. This includes an overscan correction that removes the average signal introduced by reading a CCD and bias subtraction which helps to de-bias a CCD image by subtracting the pixel-to-pixel structure in the read noise from the raw image. To further reduce the noise level of the preprocessed flat field images, a Gaussian filter with a kernel size of $16$ by $16$ pixels is applied to the overscan corrected image. 

\begin{figure}[tb]
\centering
\includegraphics[scale = 0.41]{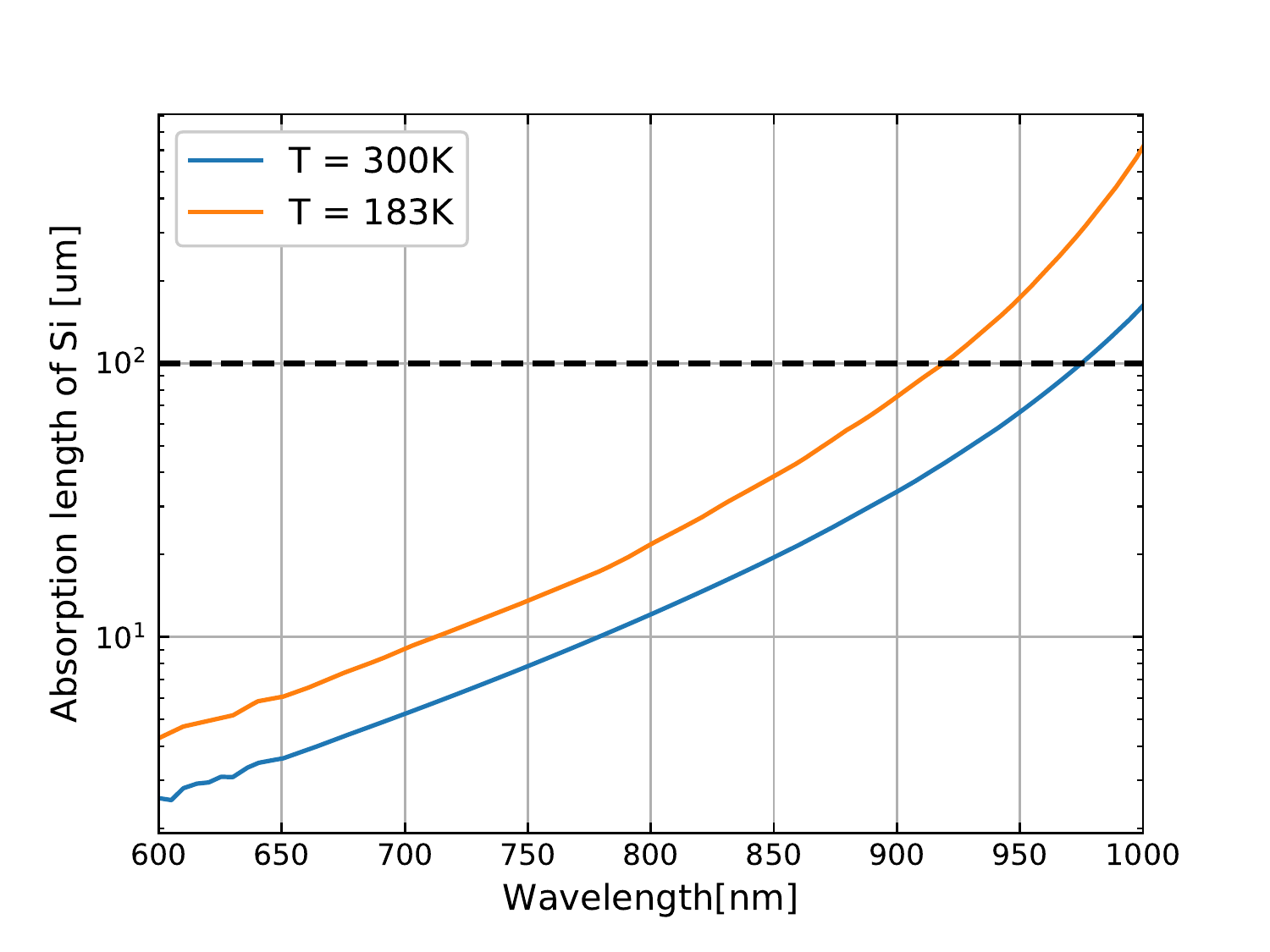}
\caption{Depth in silicon at which $99\%$ of incident light is absorbed as a function of wavelength in vacuum. {\it Blue Solid line:} Room temperature (300K). {\it Orange solid lines:} Temperature = 183K. {\it Dashed black line:} LSST CCD Silicon thickness. The refractive index of Si and temperature coefficients are adopted from \citet{Green08}. {\it Red shaded region:} wavelength range covered by the LSST y filter.} 
\label{fig:Si_ab_length}
\end{figure}
Figure~\ref{fig:Si_ab_length} shows the depth in silicon at which $99\%$ of the incident light is expected to be absorbed, as a function of wavelength under two different temperatures. For LSST CCDs with silicon thickness of $100\ \mu m$ operating under $-90 \degree C$, the silicon will start becoming transparent to incident light at $800\ \mu m$. At this point, light will reach the epoxy layer below the $100\ \mu m$ silicon and some will be reflected back to interfere with incident light in previous layers. Fringing will become more apparent at longer wavelengths as more light reaches the back of the sensor. Based on this conclusion and combined with visual inspection of the test data, fringe flat field data ranging from $880\ nm$ to $990\ nm$ in steps of $10\ nm$ are chosen to fit the fringing amplitude and derive the value of epoxy layer thickness. The fringing amplitude at an individual pixel is defined as the number of counts per pixel over the overall mean number of counts in the image with unity subtracted:

\begin{equation}
    \mbox{Fringe\ Amp.} = \frac{\mbox{Counts}}{\mbox{Overall\ Mean}}- 1
\end{equation}
\begin{figure}[tb]
\centering
\includegraphics[scale = 0.45]{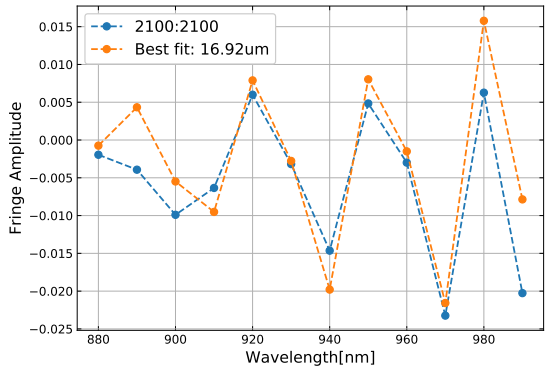}
\caption{Fringing amplitude versus wavelength at the center of e2v-CCD250-321 (pixel 2100,2100) and the best fit fringing model. {\it Orange, dashed line:} Fringing amplitude at sampled wavelength at pixel (2100,2100). {\it Blue, dashed line:} best fit fringing model for this pixel.} 
\label{fig:Fitting_exmaple}
\end{figure}

\begin{figure*}[hbt]
\centering
\includegraphics[scale = 0.6]{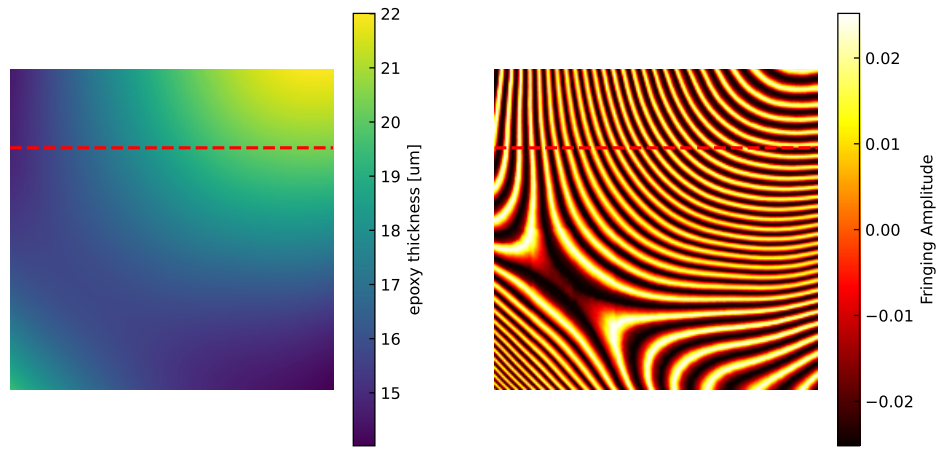}
\caption{{\it Left panel:} Derived epoxy thickness map (4k x 4k pixels) of e2v-CCD250-321.Color bar shows the range of the value of thickness. {\it Right panel:} Simulated fringing pattern at $960nm$ based on the derived epoxy thickness map for e2v-CCD250-321. A $2nm$ illumination bandwidth is assumed in simulation. Color bar shows the simulated fringing amplitude.}
\label{fig:321-map-simulation}
\end{figure*}

\subsection{Pixel by Pixel Fitting Algorithm}
To infer the thickness variation map of the epoxy layer across an entire sensor, we adopt the fitting method described in~\citet{Malumuth03}. Since only the variation of $d_{\mathrm{epoxy}}$ determines the frequency of fringes, the thickness of all the other layers are assumed to be known and constant across sensor. The boundary between each layer is assumed to be planar for simplicity (Table \ref{tbl:e2v Structure}). All the calculations are based on the assumption of colliminated beam and $2\ nm$ illumination bandwidth.
The algorithm contains the following steps: \\ \\
\textbf{Step 1:} An arbitrary pixel (we pick pixel X = 2100, Y = 2100 in the case of e2v-CCD250-321) is chosen as the starting pixel. Then simulated fringing amplitude from $880\ nm$ to $990\ nm$ is calculated for a range of epoxy thickness $d_\mathrm{epoxy}$ ranging from $5\ \mu m$ to $30\ \mu m$. The value of $d_{\mathrm{epoxy}}$ that minimizes the reduced $\chi^2$ of the fit to the observed fringing amplitude is chosen as the best fit, $d_0$, for this starting pixel. The reduced $\chi^2$ is defined as ${\chi^2}$/${(n_d-n_p)}$, with $n_d$ and $n_p$ being the number of data points and number of fitting parameters respectively. Figure~\ref{fig:Fitting_exmaple} presents the fitting results for this pixel.\\
\textbf{Step 2:} We then move to the next pixel in the same column (X = 2100, Y = 2101). Using the derived thickness value of the initial pixel, $d_0$, as a reference point, we calculate the fringing amplitudes for a set of $d_{\mathrm{epoxy}}$ values within the range of one order of fringe, $30\ nm$, centered on that value ($d_0\pm 15\ nm$). The order of fringes can be related to varying thickness of epoxy layer ($\Delta d$) via~\citep{James87}:

\begin{equation} \label{eq:thickness2forder}
    \Delta d = \frac{\delta \lambda \cos{\theta}}{4\pi n_{\mathrm{epoxy}}}
\end{equation}
In the case of normal incidence, one order of fringe ($\delta = 2\pi$) corresponds to approximately $\Delta d_{\mathrm{epoxy}} \approx 30\ nm$.
Within this given range, the value of $d_{\mathrm{epoxy}}$ minimizing the reduced $\chi^2$ of the fit is assumed to be the thickness of this pixel. And this value will be updated as the reference point for the next pixel that the algorithm will be working on. Looking for potential best fit values of $d_{\mathrm{epoxy}}$ in a limited range in this way helps to ensure the derivation of a smooth thickness variation map of the epoxy layer.\\
\textbf{Step 3:} Step 2 is repeated until reaching the end of the column (X = 2100, Y = 4000). Then we move down to the next column in that same row (X = 2101, Y = 4000), and work up the row in the same manner as described in previous steps. \\
\textbf{Step 4:} The above steps are repeated until reaching the end (X = 4000, Y = 4000). Upon this point, we move back to the initial pixel (X = 2100, Y = 2100) and repeat the same process in reverse order until reaching pixel X = 1, Y = 1.

\subsection{Fitting Results} \label{sec:Fitting Results}
Figure~\ref{fig:321-map-simulation} shows the derived epoxy layer thickness map of e2v-CCD250-321 in the left panel. Using the fringing model, we successfully reproduce the observed fringing pattern, which is shown in the right panel in Figure~\ref{fig:321-map-simulation}, based on this derived height variation map.
\begin{figure}[hbt]
\centering
\includegraphics[scale = 0.4]{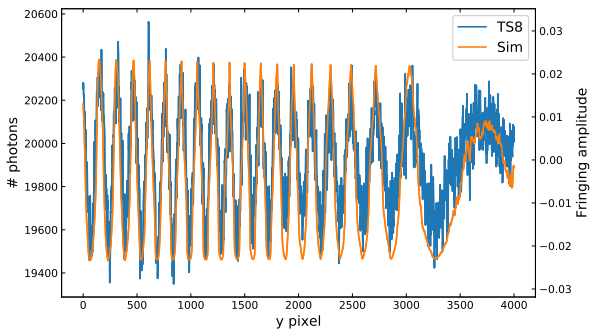}
\caption{Comparison between data and simulation for e2v-CCD250-321 column 3000, as highlighted by the red, dashed line in Figure~\ref{fig:321-map-simulation}. {\it Blue solid line:} Smoothed SLAC-TS8 flat field data. {\it Orange solid line:} simulated results under the assumption of 2nm illumination bandwidth.}
\label{fig:321-detail-compare}
\end{figure}
Figure~\ref{fig:321-detail-compare} shows the comparison between Gaussian-smoothed real data and simulation results for a specific row of the sensor. It is clear that the phases and amplitudes are in agreement between simulation and real data.

\begin{figure*}[tbh]
\centering
\includegraphics[scale = 0.45]{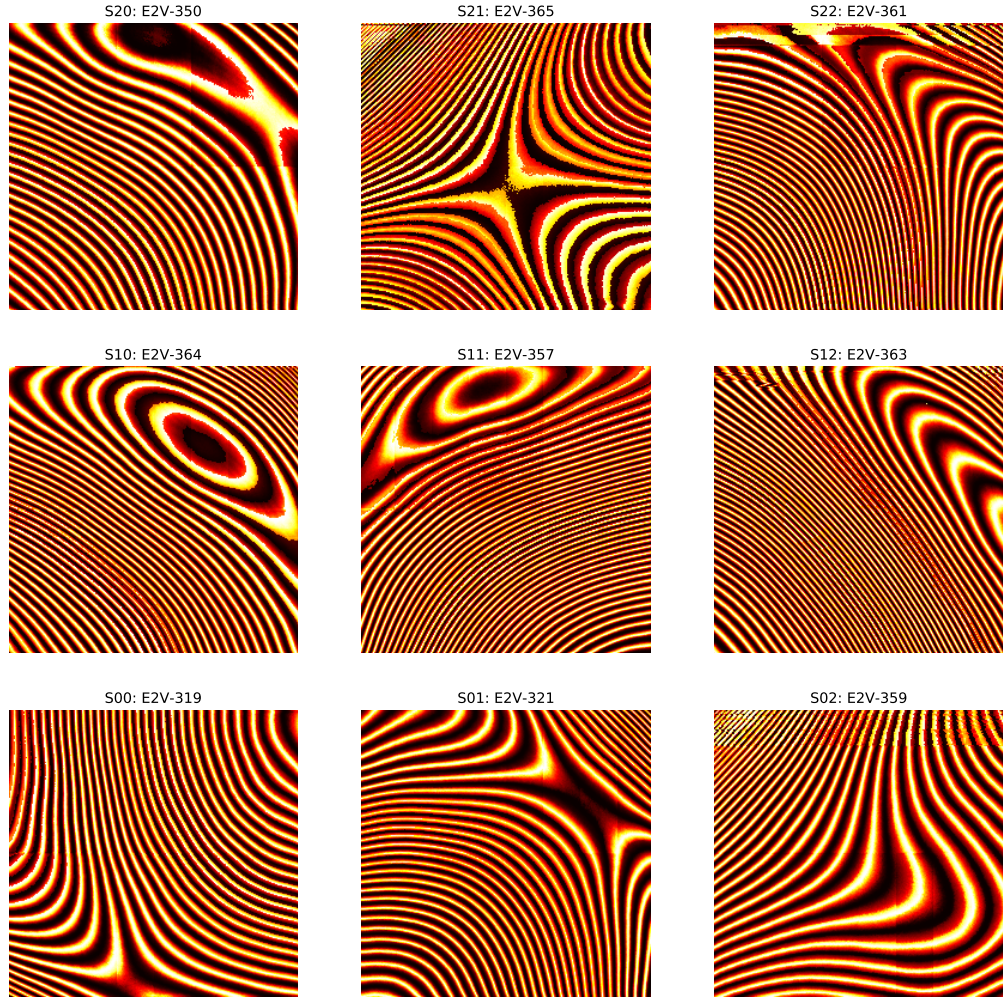}
\caption{Simulated fringing pattern at wavelength $\lambda = 960nm$ based on derived epoxy thickness map for nine e2v CCDs in RTM-020. Color scales are the same for all plots.}
\label{fig:RTM-020-SIMS}
\end{figure*}

Using this method, we generate the fringing patterns observed in the other 8 sensors in the same RTM from simulation. Figure~\ref{fig:RTM-020-SIMS} shows the simulated fringing patterns for all the nine CCDs. Most of the fringing patterns have been successfully recovered.

\section{Recipes for Realistic fringing simulation in sky background image} \label{sec:recipes for real image}
With knowledge of the compositions of sensor structure and height variation in the epoxy, we can further use the fringing model to predict the expected level of fringing in LSST images. In this section, we discuss all the ingredients needed to simulate fringing in real sky images captured by a telescope in general.  

\subsection{Hydroxyl Radical (OH) emission lines}\label{sec: OH line + y filter}
 The night sky spectrum is dominated by the emission lines produced by the rotational and vibrational transitions of hydroxyl (OH) radicals. Each vibrational transition produces a band in the observed spectrum and the transition between rotational levels associated with the two vibrational levels give rise to the fine structure of the band. The vibration-rotation spectrum of the hydroxyl radical was first observed by \citet{Meinel50a,Meinel50b}.
\begin{figure*}[htb]
\epsscale{0.85}
\centering
\plotone{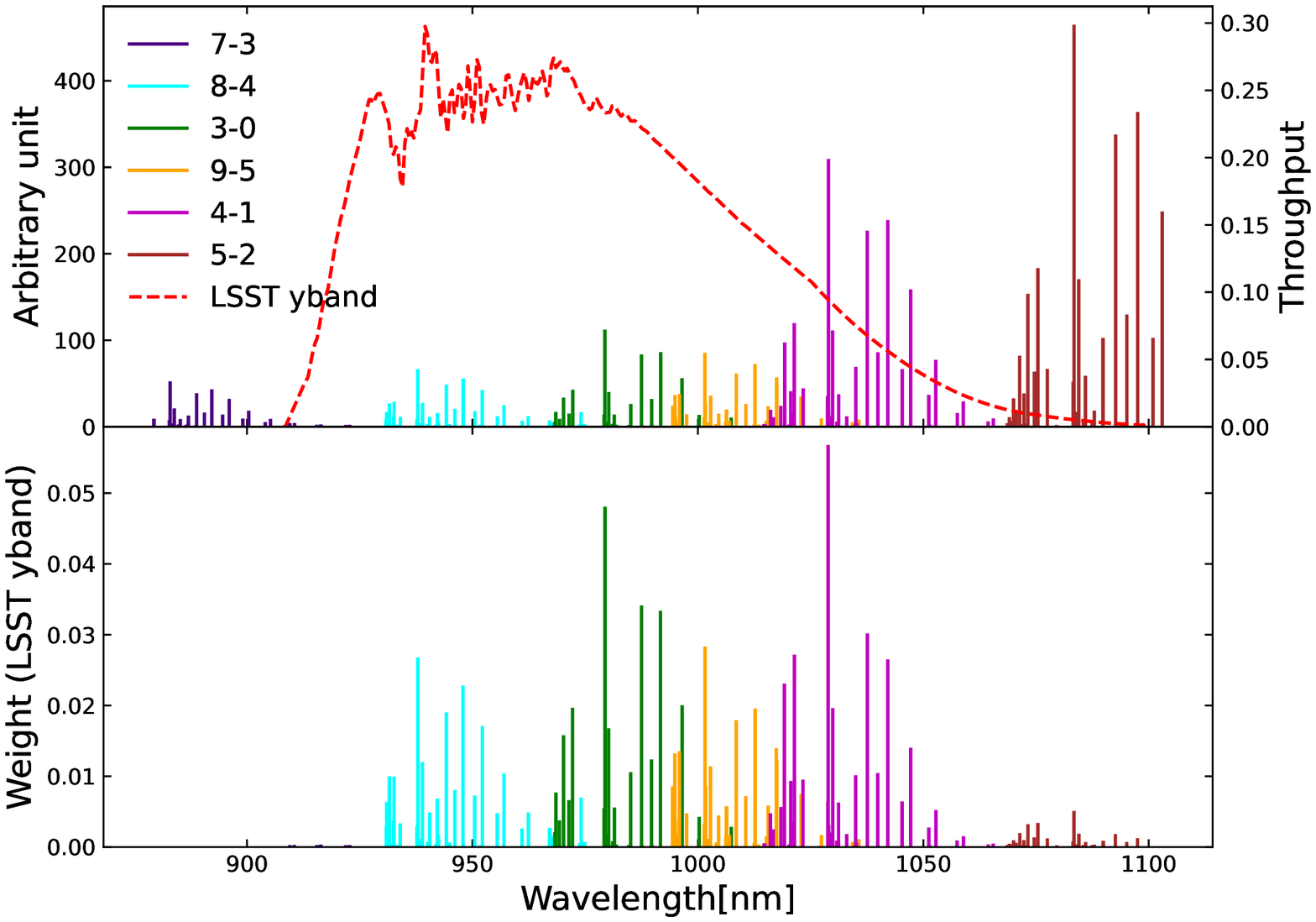}
\caption{\textbf{Top panel:} {\it Light blue line:} OH emission line from the Prime Focus Spectrograph (PFS) of the Subaru telescope~\citep{Tamura16}. {\it Red dashed line:} LSST y filter throughput curve. 
\textbf{Bottom panel:} {\it Blue line:} Normalized output of OH line intensity as weighted by LSST y filter throughput.}
\label{fig:OH_spec}
\end{figure*}
These narrow emission lines are the main sources that give rise to fringing in the observed images. The intensity of vibrational bands and the population of rotational levels within each band can be well described by Boltzmann distributions specified by vibrational temperatures, $T_\mathrm{vib}$ and rotational temperatures, $T_\mathrm{rot}$. The typical values of $T_{\mathrm{vib}}$ and $T_{\mathrm{rot}}$ are about $10000\ K$ and $200\ K$ respectively~\citep{Rousselot00}. These lines are subject to both temporal and spatial variations.  Nevertheless, the relative intensities of rotational transition lines are expected to remain roughly the same since $T_{\mathrm{tor}}$ varies much less than $T_{\mathrm{vib}}$~\citep{Noll15,Hart19}.

As discussed in Section~\ref{subsec: data reduction}, fringing will start becoming prominent in e2v sensors as the wavelength goes beyond $880\ nm$. Real images will be taken with telescope filters, and the wavelength range most relevant to fringing falls within the bandpass of the LSST y filter. The top panel in Figure~\ref{fig:OH_spec} shows the throughput curve of LSST y-band filter and the OH emission spectra taken by the Prime Focus Spectrograph (PFS) (Robert H. Lupton, private communication) of the Subaru telescope~\citep{Tamura16}. As confirmed by observations and theoretical calculations in previous studies, there are six transitions between vibrational bands ($7\rightarrow 3$, $8\rightarrow4$, $3\rightarrow0$, $9\rightarrow5$, $4\rightarrow1$, $5\rightarrow2$)~\citep{Noll15,Osterbrock96,Osterbrock97,Rousselot00} that fall within the LSST y-band. The color coding of the lines in Figure~\ref{fig:OH_spec} represents the vibrational group a line belongs to. 


When simulating fringing in the presence of OH lines, we assigned a normalized weight, $w_{\mathrm{line}}$, to each individual line. This weight is calculated based on the combination of each line's relative intensity and the value of the LSST y filter throughput curve
at corresponding wavelength. 
The bottom panel in Figure~\ref{fig:OH_spec} shows the normalized weight assigned to each line. Due to the low throughput of the LSST y filter at wavelength bluer than $900\ nm$ and redder than $1050\ nm$, the contributions to fringing from lines in vibrational groups $7-3$ and $5-2$ is small compared to those from other groups. The final simulation result, $S_{\mathrm{total}}$, is obtained by coadding the simulation image for each individual line, $S_{\mathrm{line}}$, together. This coaddition is performed as a weighted average:

\begin{equation} \label{eq:OH lines}
    S_{\mathrm{total}} = \sum_{\mathrm{line}}w_{\mathrm{line}} S_{\mathrm{line}}
\end{equation}

\begin{figure}[t]
\centering
\includegraphics[scale = 0.4]{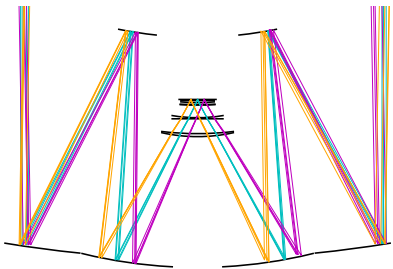}
\caption{Light path inside the LSST telescope calculated using Batoid \citep{Mayers19}.}
\label{fig:batoid-light-tracing}
\end{figure}

\subsection{Telescope optics} \label{sec:LSST optic}

\begin{figure}[t]
\centering
\includegraphics[scale = 0.55]{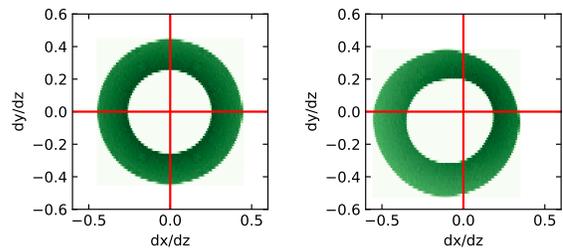}
\caption{{\it Left panel:} Incident slope distribution of LSST beam landed on the center of the focal plane. {\it Right panel:} Incident slope distribution close to the edge of the focal plane.}
\label{fig:batoid-angle-dist}
\end{figure}

\begin{figure*}[thb]
\epsscale{0.9}
\centering
\plottwo{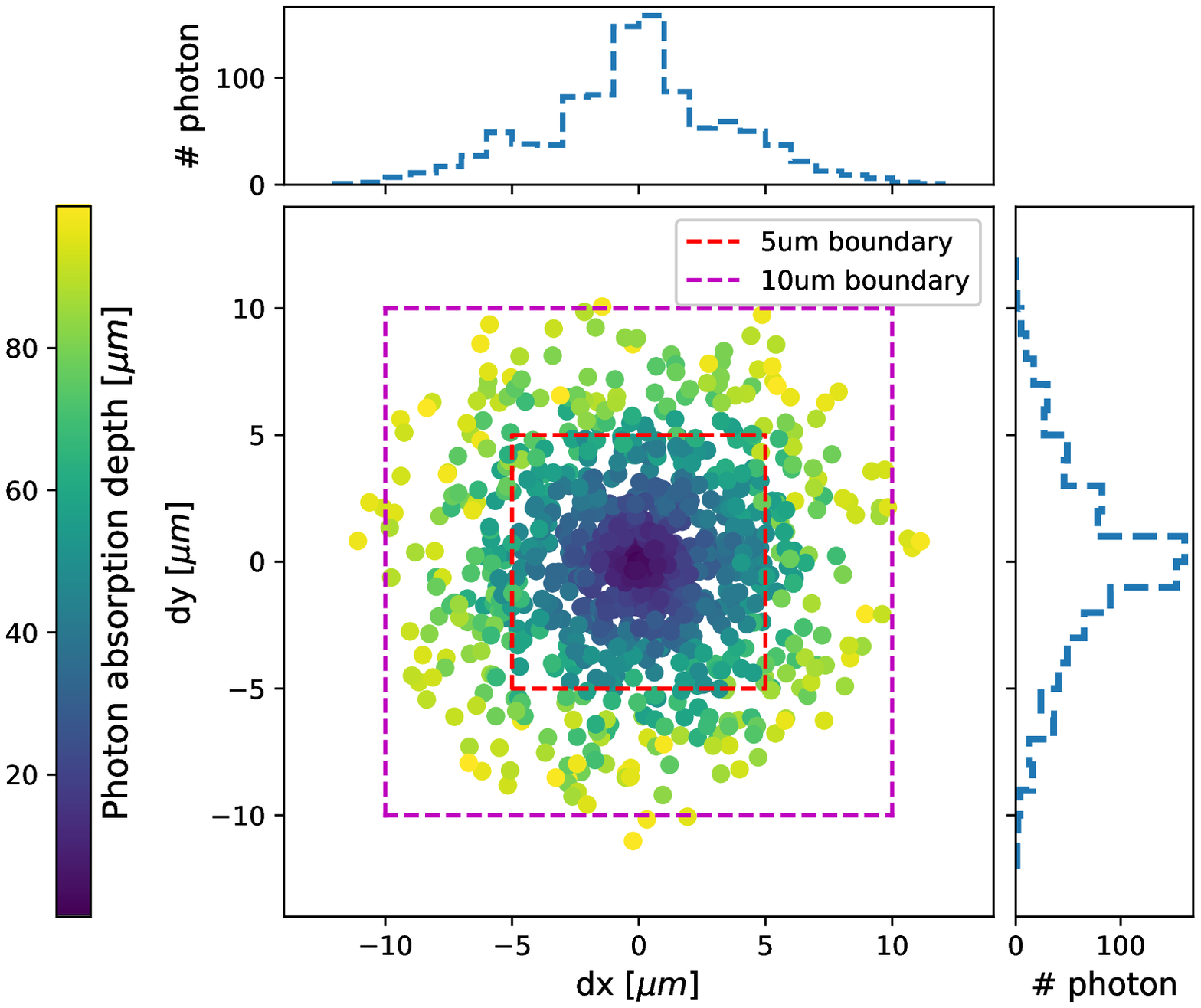}{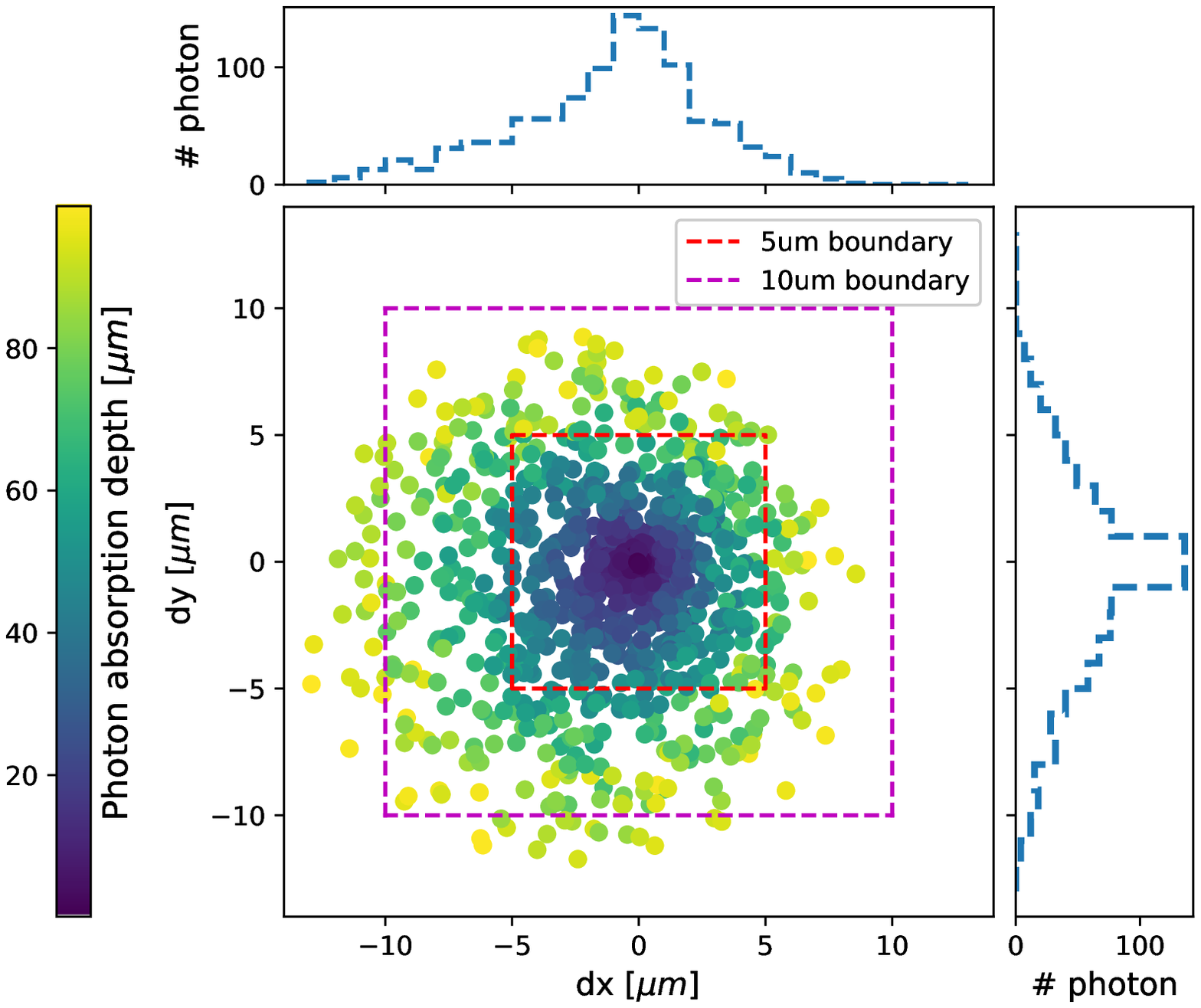}
\caption{Distribution of travelling distance in Silicon of absorbed photons in two different places of the LSST focal plane. The $5\ \mu m$ boundary shows the area of one pixel.  {\it Left plot:} in the center of the focal plane. {\it Right plot:} close to the edge of the focal plane. The colorbar indicates $z$ position (depth in silicon a photon has travelled) of absorbed photons.}
\label{fig:ab_pho_dist}
\end{figure*}

A telescope has a finite aperture that makes the incident light come from a range of angles rather than solely at normal incidence. Thus the previous assumption of collimated beams used in simulating lab results does not hold. Light arriving at different angles tends to average out the observed fringing amplitude~\citep{Groom17}. The Vera C. Rubin Observatory telescope is a three-mirror design. The three aspheric mirrors, an $8.4\ m$ primary mirror, a $3.4\ m$ convex secondary mirror, and a $5.0\ m$ tertiary mirror, give an overall focal ratio of $f/1.234$ \citep{Bond18,Ivezi19,Olivier08}. To accurately count the range of light incident angle from the $f/1.234$ beam on different positions of the LSST focal plane in fringing simulation, we employ \verb|Batoid|. \verb|Batoid| is a C++ based python optical raytracer package that characterizes the optical performance of survey telescopes based on geometric optics developed by \citet{Mayers19}. Figure~\ref{fig:batoid-light-tracing} demonstrates several examples of light paths inside the LSST telescope generated by \verb|Batoid|. In \verb|Batoid|, the directions of the incident beams are described by incoming slopes in two directions $\frac{dx}{dz}$ and $\frac{dy}{dz}$. The incident angle $\theta$ on the incidence plane, one of the inputs of the fringing model, can be derived from the two slopes: 

\begin{equation*}
    \theta (\mathrm{rad}) = \arctan{\sqrt{\left(\frac{dx}{dz}\right)^2+\left(\frac{dy}{dz}\right)^2}}
\end{equation*}
Figure~\ref{fig:batoid-angle-dist} shows the incident slope distributions of light landing on CCDs in two different locations on the LSST focal plane, one with the sensor being located in the center and the other being close to the edge of the LSST focal plane.

The incident inclination of the LSST beam makes it possible for absorbed photons in photo-sensitive region to travel into neighbouring pixels instead of staying in the same pixel where it initially landed on as in the case of normal incidence. To check if this may affect the fringing simulation results in a significant way, we did a detailed simulation in which each absorbed photon is tracked to its final location where the electron-photon pair is generated in the photo-sensitive region of CCD. Figure~\ref{fig:ab_pho_dist} presents the distribution of the distances photons have travelled in x and y directions as specified by the two slopes before getting absorbed in Silicon for 1000 photons assumed to be landed at the center of a pixel.  The two slopes will also change upon refraction into silicon. The two cases presented in Figure~\ref{fig:ab_pho_dist} correspond to the angle distributions showed in Figure~\ref{fig:batoid-angle-dist}. The colorbar indicates the z direction (depth) that a photon has travelled in Silicon. Each pixel of LSST CCD sensor is $10\ \mu m$ in width and length, and $100\ \mu m$ in depth~\citep{Ivezi19}. Simulation results in Figure~\ref{fig:ab_pho_dist} indicate that inter-pixel migration of absorbed photons do exist. However, even in extreme case where light lands on the focal plane edge, $98\%$ of the photons will travel less than $10 \mu m$, which corresponds to the size of one pixel. Since the typical size of observed fringe is about $20-30$ pixels, widening by one pixel will merely affect the simulation result. Thus we consider this effect to be negligible for the purpose of simulating the fringing pattern of 4000 by 4000 pixels images. Thus, to save computational time, an absorbed photon is always assumed to end up in the same pixel as the one it initially landed on. Additionally, the distribution of incident slopes of light is assumed to be constant for all pixels of a CCD since this distribution is expected to vary very slowly across the focal plane.

To account for the range of incident angles, the fringing simulation at given wavelength is obtained via a weighted average of simulations over all the angles. We can further write Eq~\ref{eq:OH lines} as:

\begin{equation} \label{eq:sim_equation}
    S_{\mathrm{total}} = \sum_{\mathrm{line}}w_{\mathrm{line}} S_{\mathrm{line}} = \sum_{\mathrm{line}}w_{\mathrm{line}}\sum_{\theta}w_{\theta}S_{\theta}
\end{equation}
where $S_\theta$ is an individual fringing simulation for a given incident angle $\theta$ with normalized weight $w_\theta$ derived based on the angle distributions.  Eq~\ref{eq:sim_equation} implies that simulating fringing in sky background images with real telescope optics properly requires coadding simulations over all the OH lines and over all the light incident angles.

\begin{figure*}[thb] 

\epsscale{1}
\centering
\plottwo{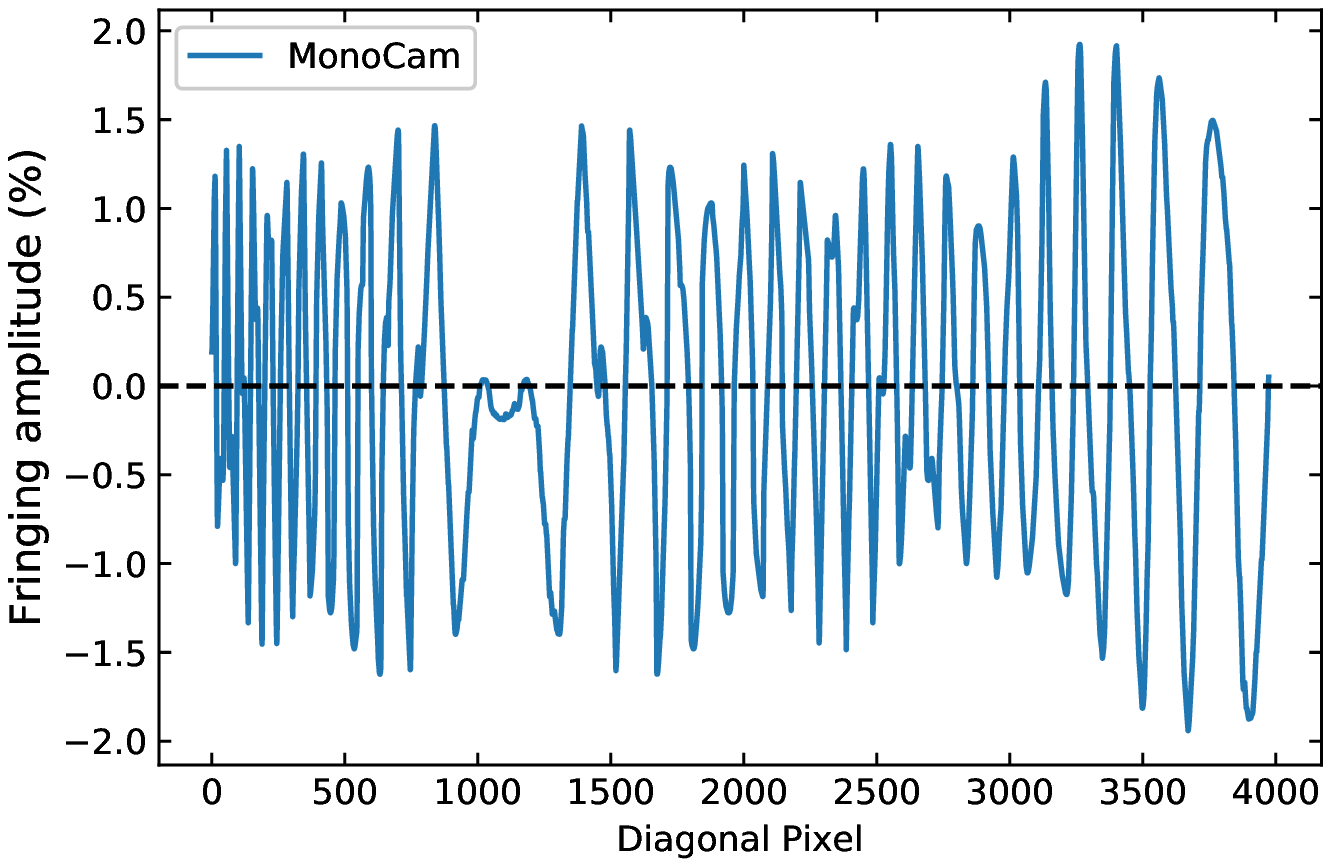}{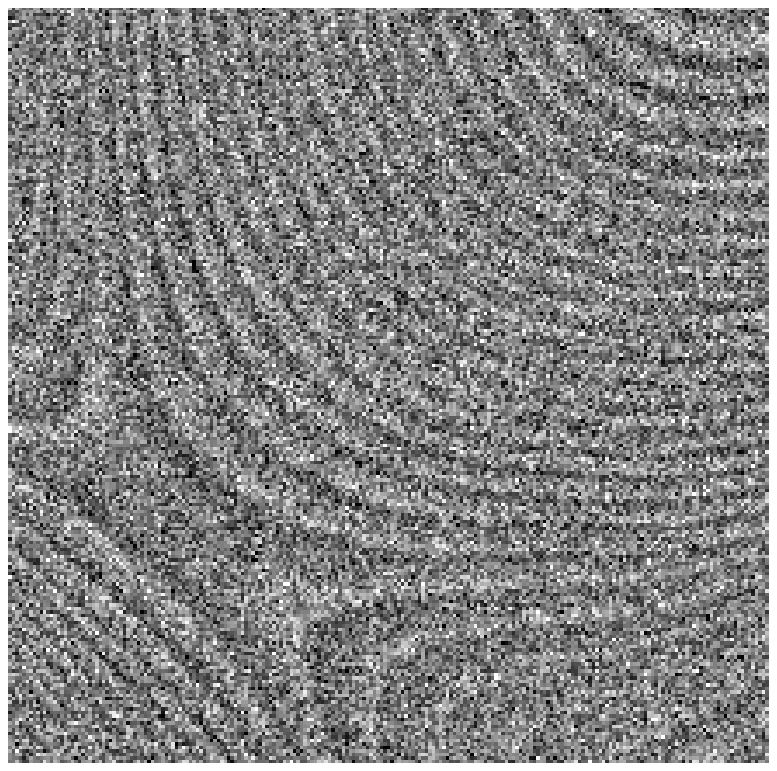}
\caption{{\it Left plot:} Simulated fringing amplitude for the diagonal pixels of e2v-CCD250-321 based on OH emission lines, MonoCam optics and LSST y filter. {\it Right plot:} 4kx4k simulated MonoCam midnight sky background image at 1.2 airmass with 60-second Poisson photon noise and Gaussian read noise added.}
\label{fig:MonoCam_sims}
\end{figure*}

\section{Fringing simulation results} \label{sec:real_sky}
To verify the robustness of the fringing model in simulating real sky images, we first apply it to simulating sky background fringing of MonoCam \citep{Brooks17} and Hyper-Suprime Camera (HSC) \citep{Miyazaki18} before making predictions for LSST. In this section, we first discuss the comparison between simulation results and observations in terms of the optics setup of MonoCam and HSC. Then we use the fringing model to predict the expected level of fringing in LSST sky background images. All simulations in this section follow the methods described in Section~\ref{sec:recipes for real image}.

\subsection{Fringing of MonoCam} \label{sec: MonoCam sim}

\begin{figure*}[bt]
\centering
\includegraphics[scale = 0.5]{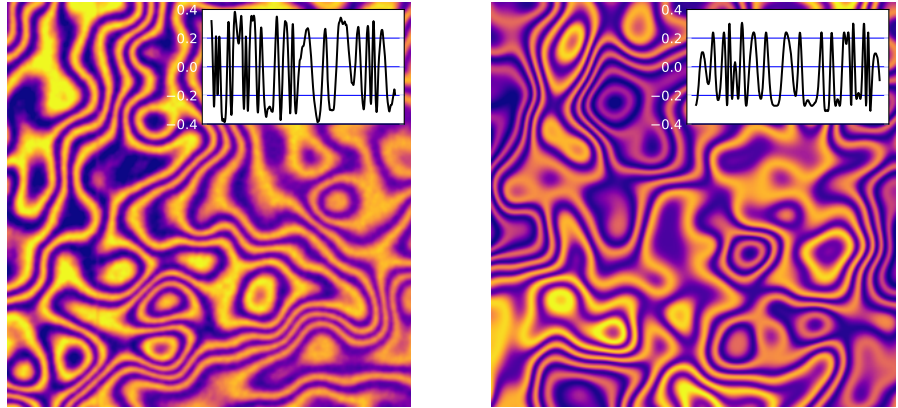}
\caption{Fringing in HSC CCD. {\it Left panel}: Fringing observed in a 1000 by 1000 pixel region of a single HSC CCD sensor. {\it Right panel}: Simulated, noiseless HSC fringing pattern based on OH emission lines, HSC optics and HSC y filter throughput. The inset figures in the top right of each figure shows the fringing amplitude ($\%$) along the diagonal pixels.}
\label{fig:HSC_sims}
\end{figure*}

MonoCam was a camera employing a single LSST prototype e2v-CCD250 sensor, with reported fringing amplitude of around $2\%$. The data were taken with a 1.3m reflector telescope with a overall focal ratio of $f/4.0$ and with LSST y filter under a temperature of $-120\degree C$ \citep{Brooks17}. Since the CCD used in MonoCam came from the same manufacturer as the sensor studied in this paper, the derived epoxy thickness map of e2v-321 is assumed for this prototype sensor for the purpose of fringing simulation. 

The left panel of Figure~\ref{fig:MonoCam_sims} presents the fringing amplitude of the diagonal pixels of the simulated, noiseless image. The simulation result shows that the MonoCam fringing amplitude is about $1.5\%$, which agrees to the amplitude of the smoothed and noise-reduced midnight fringing pattern given in Figure 9 of~\citet{Brooks17}. A full sensor image of the simulated sky background is shown in the right panel of Figure~\ref{fig:MonoCam_sims} based on the $60$ second exposure time~\citep{Brooks17} of MonoCam. Poisson photon noise and Gaussian read noise of CCD are added to the simulated full image by using the \verb|Galsim| module~\citep{Galsim}. It is clear that a nontrivial fringing pattern still appears in the image. These results suggest that our fringing model simulation is in good agreement with MonoCam observation.

\subsection{Fringing in HSC CCD sensor}

HSC is an $870$ megapixel prime focus optical imaging camera with a overall focal ratio of $f/2.0$ implemented on the $8.2\ m$ Subaru telescope. $116$ fully-depleted  2048 × 4096 pixel CCDs with a thickness of $200\ \mu m$ are employed in the focal plane \citep{Miyazaki18}. The HSC optics offers an opportunity to test the fringing model's response to a fast input beam. From the inspection of real HSC sky images, we find that fringing of HSC CCD has a sensor-dependent amplitude ranging from $0.2\%$ to $0.6\%$. As an example, the left panel of Figure~\ref{fig:HSC_sims} shows the observed fringing pattern in a 1000x1000 pixel region of an HSC CCD. To reduce the noise and make fringes easier to see, the plotted data have been smoothed by a 16x16 pixel Gaussian kernel. The mini panel in the top right of the figure shows the fringing amplitude, which is about $0.3-0.4\%$, along the diagonal pixel of the image. Fringes in HSC CCDs are likely to be caused by the non-uniformity in the $200\ \mu m$ silicon layer. This is because they only exhibit one set of fringes with similar features as the ones observed in other back-illuminated sensors, such as the HRC CCD and WFC CCD, whose fringes are modelled based on height variation in silicon detection layer, studied by \citet{Walsh03}. 

To simulate HSC sensor fringing, we made the following changes to the sensor model as depicted in Table~\ref{tbl:e2v Structure}. First, the thickness of epoxy is kept constant at $14\ \mu m$ since HSC CCD fringing is caused by non-uniformity in 200$\ \mu m$ silicon instead of epoxy as discussed above. Second, a Guassian Random Field (GRF) is used to characterize the variation of the detection layer. Since our goal to verify that the fringing model can produce comparable level of fringing amplitude as the one measured from a real HSC image, the actual fringing pattern can be arbitrary. Thus, GRF is a good approximation to the underlying thickness change of the silicon layer, whose variation can be inferred by counting the number of observed fringes across the image. Using Eq~\ref{eq:thickness2forder} and plugging in values for silicon ($n_{\mathrm{si}} = 3.6$, $\lambda = 1000\ nm$), we know that a $0.139\ \mu m$ change in silicon thickness will give one order of fringe in HSC CCDs. A visual inspection on the HSC fringing image (left panel of Figure~\ref{fig:HSC_sims}) suggests that there are $10$ to $15$ fringes across the image. This implies the overall variation of this particular patch is about $1.39\ \mu m$ to $2.08\ \mu m$. These values set the scale of the GRF used to characterize the silicon layer of HSC CCD. We choose to use the average value, $1.74\ \mu m$, as the height variation of the GRF ($[199.13\ \mu m, 200.87\ \mu m]$) across 1000 by 1000 pixel.
\verb|Batoid| is then used to generate the angle distribution of the incident beam in the center of the focal plane based on HSC telescope optics. The OH line intensities are weighted by the throughput curve of the HSC y-filter following the discussion of Section~\ref{sec: OH line + y filter}. In the right panel of Figure~\ref{fig:HSC_sims}, we present the simulated, noiseless 2D fringing pattern and amplitude across the diagonal pixels of simulated image. It is clear that the predicted  amplitude ($\sim 0.3\%$) is comparable to the observed level in the left panel.

\subsection{Prediction of Fringing in LSST sky background images}
\begin{figure}[htb]
\centering
\includegraphics[scale = 0.41]{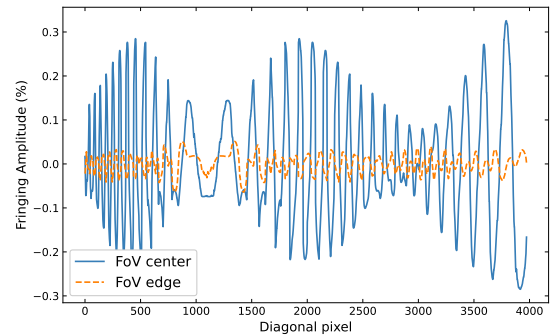}
\caption{Simulated, noiseless fringing pattern at different locations on the LSST focal plane for diagonal pixels of e2v-CCD250-321 based on LSST optics setup. {\it Blue, solid line:} sensor at the center. {\it Orange, dashed line:} sensor being at the edge of the focal plane. The corresponding angle distributions of incident beam for the two cases are given in Figure~\ref{fig:batoid-angle-dist}.}
\label{fig:LSST_sims}
\end{figure}


\begin{figure*}[tbh]
\centering
\includegraphics[scale = 0.42]{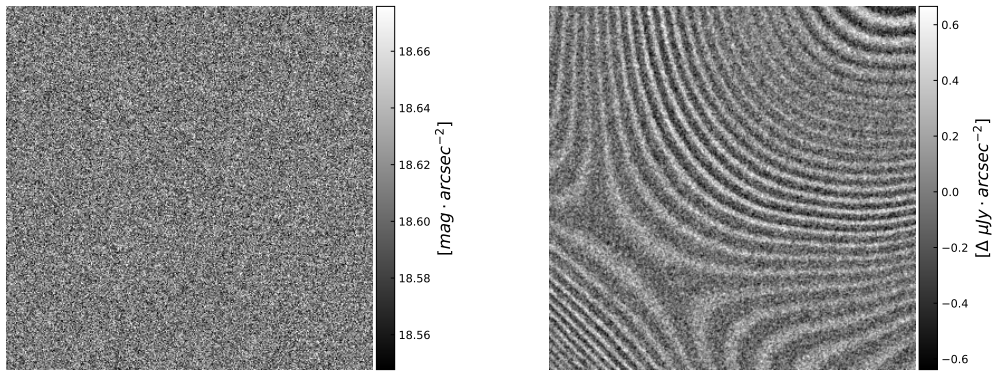}
\caption{{\it Left panel:} Simulated, single exposure sky background image in LSST y band. Colorbar shows the sky brightness in unit of mag arcsec$^{-2}$. {\it Right panel:} A Gaussian filter with kernel size of 16 by 16 pixel applied to the simulated image.  The color bar shows the deviation from the mean value in $\mu$Jy arcsec$^{-2}$.}
\label{fig:lsst-image}
\end{figure*}

After validating the fringing model with MonoCam and HSC optics, we simulated the expected fringing pattern of e2v-CCD250-321 based on LSST setups as described in Section~\ref{sec:LSST optic}.  In Figure~\ref{fig:LSST_sims}, we compared the simulated fringing patterns for two cases in terms of a CCD's location on the LSST focal plane, which differ in their angle distribution of incident beam as shown in Figure~\ref{fig:batoid-angle-dist}. Compared with the values calculated in previous sections, the fringing amplitude for LSST decreased to $0.2\%$. This is caused by the fact that the wide range of incident angle of LSST optics decreases the overall fringing amplitude when incorporating the beams coming from all the angles in the simulation. As the sensor is placed furhter away from the center of field of view, the fringing amplitude decreases to $0.04\%$. This is because the light arriving at the edge of the focal plane does not come from an exact $f/1.23$ beam due to the LSST optical design~\citep{Olivier08}, as manifested in Figure~\ref{fig:batoid-angle-dist}. This leads to a wider range of incident angle, which further decreases the overall fringing amplitude when incorporating the beams coming from all the angles in the simulation. 

The left panel of Figure~\ref{fig:lsst-image} shows a simulated LSST sky background image with Poisson photon noise and Gaussian read noise added for a single exposure of $30$ second. The values have been converted from counts per pixel in the original simulation to surface brightness unit, mag arcsec$^{-2}$, by properly accounting for the sensor gain, yband zero point and pixel scale. The sky brightness level in our simulated image is close to both the expected value\footnote{\url{https://smtn-002.lsst.io/}} and in-situ measurement of the sky level in LSST y-band~\citep{High10}. Due to the low amplitude of fringing and relatively short single exposure time, fringing can not be easily observed as the image is mostly dominated by noise. However, after applying a 2D Gaussian kernel, which serves as a low pass filter to extract low amplitude structure on relatively large scales in image progressing, to the image, the fringing patterns become apparent as shown in right panel of Figure~\ref{fig:lsst-image}. We discuss its implication in detail in the next section. 

\subsection{Impact of LSST fringing on measurements}



During LSST survey operation, images taken by the Rubin observatory will be processed by Rubin's LSST Data Management Science Pipelines software stack\footnote{\url{https://pipelines.lsst.io/}}, developed by Rubin's LSST Data Management (DM) team. We refer readers to ~\citet{Bos18,Bosch19} for more a detailed description of the the LSST image processing pipeline. To better quantify the impact of fringing in the context of data measurement and image reduction, it is more convenient for us to convert units to~\ujyac. The overall variation caused by fringing across the image, as shown by the colorbar of the right panel of Figure~\ref{fig:lsst-image}, is about 0.6~\ujyac.
Based on the image post-processing results from HSC~\citep{Aihara19,Aihara22}, which uses a custom pipeline, known as \verb|hscPipe|, that has been built upon the LSST Data Management codebase, the current measurement has a $10\%$ error at the limiting surface brightness level of 26th AB magnitude, which approximates to $0.015$~\ujyac. The predicted level of variation from fringing is already $40$ times larger than this value. In the future, with careful post-analysis, the surface brightness limit for LSST is expected to be 30th AB magnitude, which further reduces the $10\%$ measurement error to $0.004$~\ujyac (Robert H. Lupton, private communication). Thus, it is clear that fringing will impact measurements on single exposure images.

Principal Components Analysis (PCA) method has been proved to be an effective approach for fringing correction~\citep{Waters20,Medford21}. In the PCA method, a set of orthogonal images/components are trained from a sample of fringing images. Linear combinations of these components can be used to construct bias fringe images, which will then be subtracted from the target image to remove fringing. However, since sky emission lines will vary temporally as mentioned in Section~\ref{sec: OH line + y filter}, fringing is also expected to vary over time. 
Since real LSST images are unavailable at this time, we infer the number of components needed for LSST fringing correction for a single CCD from simulation. In this study, we use the temporal variation of relative intensities of the OH lines within LSST y-band over one night from \citet{Noll15} to generate 11 simulated sky background fringing images at each hour spanning from 7pm to 5am for training. For PCA calculation, we use code from the \verb|scikit-learn| package~\citep{scikit-learn}. We find that only two components are needed to fully characterize fringing in one CCD in this case, with the first component having $98\%$ variance. The second component, which characterizes the temporal variation of fringing, has $2\%$ variance. This implies that solely based on the one night's worth of data of OH line variation, as given in \citet{Noll15}, the fringing pattern is expected to vary at the $2\%$ level and we should be able to describe the fringes with two patterns for removal. However, the data from \citet{Noll15} might not fully represent the OH line variation at the site of Rubin Observatory. Future studies should investigate a longer term data sample to determine if the emission line data used in this study are representative enough and if night-to-night variations might be larger.

\begin{figure}[t]
\centering
\includegraphics[scale = 0.5]{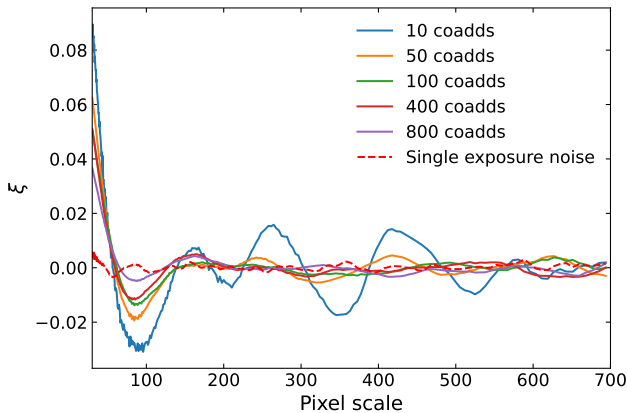}
\caption{Correlation functions between pixel , as a function of pixel scale across sensor for increasing number of simulated coadds. Red, dashed line shows the null case where there is no fringing in the simulated background image.}
\label{fig:coadd_cor}
\end{figure}

In terms of the impact of fringing in coadded images, it will be hard to properly characterize it until the effectiveness of the fringing removal algorithm of the Rubin's LSST DM pipeline is known. Here, we study the effect of fringing in coadds for the case in which no fringing correction has been made in single exposure images from random sensors. This is done by looking at correlations between pixels at different scales for increasing number of coadds. The coadds are simulated via stacking regions of randomly rotated single exposure sky background images, as picked randomly from the 9 sensors in Figure~\ref{fig:RTM-020-SIMS}, together. Figure~\ref{fig:coadd_cor} shows the correlation function of the mean subtracted pixel fluxes as a function of pixel scale in terms of increasing number of coadds. We use \verb|TreeCorr|~\citep{Jarvis04} to calculate the correlation functions. For comparison, we also did a null case in which we measured the same correlation function of a random single exposure background image without fringing. Based on Figure~\ref{fig:coadd_cor}, it is clear that even at the level of 800 coadds, there is a still difference in the correlation function when compared to that of the null case. Based on the current observation strategy of LSST~\citep{Scolnic18}, the numbers of y-band exposures of the Wide Fast Deep survey (WFD) and  Deep Drilling Field mini-survey (DDF) after the 10-year survey length of  LSST are $180$ and $2600$ respectively. This implies that, without applying any fringing correction to single exposure images, fringing will still make noticeable impact in coadded images for WFD at least. To truly characterize fringing in coadded images, we will need to quantify the effectiveness of fringing removal algorithm by running simulated fringing images through DM pipeline as a future study.
\begin{deluxetable*}{ccccccc}[hbt] 
\tablecaption{Observed and simulated fringing amplitude of e2v sensor for different optical setups}
\tablenum{2}
\tablecolumns{7}
\tablehead{
\colhead{} &
\colhead{Optics} &
\colhead{Temperature} &
\colhead{Source of fringing} &
\colhead{Simulation setup} &
\colhead{Observed amplitude} &
\colhead{Simulated amplitude}
}
\startdata
 SLAC-TS8 & $f/\infty$ & $-90\degree C$ & Epoxy & Monochromator ($960nm$) &$\sim 2\%$ & $2\%$   \\
 MonoCam & $f/4$ & $-120\degree C$ & Epoxy &OH lines $+$ LSST y band & $\sim 2\%$ &  $1.5\%$\\
 HSC & $f/2$ & $-100\degree C$ & Silicon($\sim 200\mu m$) &OH lines $+$ HSC y band &$0.2\%\sim 0.6\%$ & $0.3\%$\\
 LSST & $f/1.23$ & $-100\degree C$ &Epoxy & OH lines $+$ LSST y band & $ -$ & $ 0.04\%\sim 0.2\%$ \\
\enddata

\end{deluxetable*}
\label{tbl:fringing summary}

\section{Summary}
We have presented a fringing model for e2v CCD sensors implemented on the Rubin's LSST camera focal plane. We have demonstrated that these observed fringes in e2v CCDs from SLAC-TS8 are caused by the thickness variations of the epoxy layer that glues the sensor and the support silicon together. We have shown that this model allows us to simulate the fringing patterns accurately observed in lab data on a pixel-by-pixel level. 

We have shown that with sufficient flat field data taken with close wavelength spacing ($< 10nm$) and under certain assumptions for the illumination setups, such as illumination bandwidth and incident angle, an underlying thickness map as a function of pixel position can be derived for each sensor via the fitting algorithm adopted from previous studies. Based on the derived thickness map, we have successfully reproduced the fringing patterns observed in nine e2v sensors in one Raft Tower Module on LSST focal plane from SLAC-TS8.

We have demonstrated that, by properly incorporating all the relevant factors, such as telescope optics and OH emission lines that dominate the night sky spectra, into the fringing model, we are able to recover the observed fringing amplitude for MonoCam and HSC. We then use this model to predict the fringing pattern in LSST real sky images and find that the simulated LSST fringing amplitude ranges from $0.04\%$  to $0.2\%$ depending the location of a CCD on the focal plane. Table~\ref{tbl:fringing summary} summarizes all the simulation results in terms of different optics setups and conditions. 

Finally, we have shown that fringing will be nontrivial for Rubin's LSST. In the case of single exposure, the variation in surface brightness caused by fringing is found to be $40$ times larger than the current measurement error. By using a PCA method, we have shown that $2$ components are needed to correct for fringings in single exposure images within the time scale of a single night. And more long-term fringing images are needed to account for the temporal variation in fringing caused by changing in the sky emission lines over larger time scale. In the case of coadded images, by studying correlation functions of simulated coadd images in pixel space, we find that fringing level is still greater than the noise level even in the case of 800 coadded images, assuming no fringing removal algorithm has been applied. It is thus necessary to include fringing correction alogrithm in Rubin's LSST image processing pipeline and characterize the impact of fringing in coadded images.

\acknowledgements
This paper has undergone internal review by the LSST Dark Energy Science Collaboration. The internal reviewers were David Kirkby, Josh Meyers, and Andrew Bradshaw.

We are grateful to Pierre Antilogus, Jim Chiang, Mike Jarvis, Lee Kelvin, David Kirkby, Josh Meyers, Andrei Nomerotski, Paul O'Connor, Andy Rasmussen, Aaron Roodman and Peter Yoachim for helpful discussions.

ZG and CW were supported by Department of Energy, grant DE-SC0010007. CSL gratefully acknowledges financial support from DOE grant DE-SC0009999 and Heising-Simons Foundation grant 2015-106.

The DESC acknowledges ongoing support from the Institut National de Physique Nucl\'eaire et de Physique des Particules in France; the Science \& Technology Facilities Council in the United Kingdom; and theDepartment of Energy, the National Science Foundation, and the LSST Corporation in the United States.  DESC uses resources of the IN2P3 Computing Center (CC-IN2P3--Lyon/Villeurbanne - France) funded by the Centre National de la Recherche Scientifique; the National Energy Research Scientific Computing Center, a DOE Office of Science User Facility supported by the Office of Science of the U.S.\ Department ofEnergy under Contract No.\ DE-AC02-05CH11231; STFC DiRAC HPC Facilities, funded by UK BEIS National E-infrastructure capital grants; and the UK particle physics grid, supported by the GridPP Collaboration.  This work was performed in part under DOE Contract DE-AC02-76SF00515.

The contributions from the primary authors are as follows. ZG analysed data, built the fringing model, formulated the fitting algorithm, developed the simulation code and wrote the paper. CW supervised the project and analysis. CL conducted physical measurements and provided layer composition information of the e2v sensors. RL provided the PFS OH emission line data and advised on the characterization of the impact of fringing on single exposure images and coadds.
\newpage

\bibliography{reference}{}
\bibliographystyle{aasjournal}



\end{CJK*}
\end{document}